 \documentclass[preprint,superscriptaddress,showpacs,preprintnumbers,amsmath,amssymb]{revtex4}

\usepackage{graphicx}
\usepackage{dcolumn}
\usepackage{bm}
\usepackage{mathrsfs, booktabs, multirow}
\usepackage{CJK}
\usepackage{color}

\newcommand{\br}{\mbox{\boldmath$r$}}
\newcommand{\bj}{\mbox{\boldmath$j$}}
\newcommand{\svec}[1]{{\mbox{\boldmath$#1$}}}

\begin{document}

\title{One-pion exchange current corrections for nuclear magnetic moments in relativistic mean field theory}

\author{Jian~Li~}
 \affiliation{School of Physics, State Key Laboratory of Nuclear Physics and Technology, Peking University, Beijing 100871, China}

\author{J. M. Yao}
 \affiliation{School of Physical Science and Technology, Southwest University, Chongqing 400715, China}

\author{J. Meng}
 \email{mengj@pku.edu.cn}
 \affiliation{School of Physics, State Key Laboratory of Nuclear Physics and Technology, Peking University, Beijing 100871, China}
 \affiliation{School of Physics and Nuclear Energy Engineering, Beihang University, Beijing 100191, China}

\author{A. Arima}%
 \affiliation{Science Museum, Japan Science Foundation, Tokyo 102-0091, Japan}

\date{\today}

\begin{abstract}
The one-pion exchange current corrections to isoscalar and isovector
magnetic moments of double-closed shell nuclei plus and minus one
nucleon with $A=15,17,39$ and $41$ have been studied in the
relativistic mean field (RMF) theory and compared with previous
relativistic and non-relativistic results. It has been found that
the one-pion exchange current gives a negligible contribution to the
isoscalar magnetic moments but a significant correction to the
isovector ones. However, the one-pion exchange current doesn't
improve the description of nuclear isovector magnetic moments for
the concerned nuclei.
\end{abstract}

\pacs{21.10.Ky, 
      21.30.Fe, 
      21.60.Jz  
     }

\maketitle


Nuclear magnetic moment is one of the most important physics
observables. It provides a highly sensitive probe of the
single-particle structure, serves as a stringent test of nuclear
models, and has attracted the attention of nuclear physicists since
the early days~\cite{Blin-Stoyle1956,Arima1984}.

The static magnetic dipole moments of ground states and excited
states of lots of atomic nuclei have already been measured with
several methods~\cite{Stone2005}. With the development of the
radioactive ion beam (RIB) technique, it is now even possible to
measure the nuclear magnetic moments of many short-lived nuclei near
the proton and neutron drip lines with very high
precision~\cite{Neyens2003,Yordanov2007,Tripathi2008}.

Theoretical description for nuclear magnetic moment is a
long-standing problem. For the last decades, many successful nuclear
structure models have been built up. However, the application of
these models for nuclear magnetic moments is still not satisfactory.

The Schmidt values predicted by the extreme single-particle shell
model qualitatively succeeded in explaining the magnetic moments of
odd-$A$ nuclei near double-closed shells. Later on, the magnetic
moment of nuclei farther away from closed shells were found to be
sandwiched by the Schmidt lines. Therefore, lots of efforts have
been made to explain the deviations of the nuclear magnetic moments
from the Schmidt values. In shell model, the first-order
configuration mixing (core polarization)~\cite{Arima1954}, i.e., the
single-particle state coupled to more complicated $2p-1h$
configurations, and the second-order core polarization as well as
the meson exchange current
(MEC)~\cite{Chemtob1969,Shimizu1974,Towner1983} are taken into
account to explain the deviations.

The magnetic moments of $LS$ closed shell nuclei plus or minus one
nucleon are of particular importance, in which, there are no
spin-orbit partners on both sides of the Fermi surface and therefore
all first-order core polarization corrections vanish. It has been
shown in non-relativistic calculations that the second-order core
polarization effect dominates the deviations of isoscalar magnetic
moments and also gives large corrections to the isovector magnetic
moments~\cite{Towner1987,Arima1987}. The MEC effect, due to its
isovector nature, gives rather small corrections to the isoscalar
magnetic moments while gives important corrections to the isovector
magnetic moments~\cite{Towner1983,Ichii1987}. As a result, the
calculated corrections to the isoscalar magnetic moments are in
reasonable agreement with the data, and the net effect of second
order core polarization and MEC gives the right sign for the
correction to the Schmidt isovector magnetic
moments~\cite{Towner1987,Arima1987}.

In the past decades, the RMF theory, which can take into account the
spin-orbit coupling naturally, has been successfully applied to the
analysis of nuclear structure over the whole periodic table, from
light to superheavy nuclei with a few universal
parameters~\cite{Ring1996,Vretenar2005,Meng2006}. However, a
straightforward application of the single-particle relativistic
model, where only sigma and the time-like component of the vector
mesons were considered, cannot reproduce the experimental magnetic
moments~\cite{Miller1975,Serot1981}. It is because the reduced Dirac
effective nucleon mass ($M^*\sim0.6M$) enhances the relativistic
effect on the electromagnetic current~\cite{McNeil1986}. After the
introduction of the vertex corrections to define effective
single-particle currents in nuclei, e.g., the ``back-flow" effect in
the framework of the relativistic extension of Landau's Fermi-liquid
theory~\cite{McNeil1986} or the random phase approximation (RPA)
type summation of p-h and p-$\bar{n}$ bubbles in relativistic
Hartree approximation~\cite{Ichii1987a,Shepard1988}, or the
consideration of non-zero space-like components of vector meson in
the self-consistent deformed RMF
theory~\cite{Hofmann1988,Furnstahl1989,Yao2006}, the isoscalar
magnetic moment can be reproduced quite well. Unfortunately, these
effects cannot remove the discrepancy existing in isovector magnetic
moments. To eliminate the discrepancy, the MEC corrections have been
investigated in the linear RMF theory in Ref.~\cite{Morse1990},
which was found to be significant but enlarge the disagreement with
data.

In view of these facts, it is essential to investigate the nuclear
magnetic moments in the RMF theory with modern effective
interactions. In this work, the isoscalar and isovector magnetic
moments of light odd-mass nuclei near the double-closed shells will
be studied in axially deformed RMF theory with the consideration of
non-zero space-like components of vector meson. In particular, the
one-pion exchange current corrections to nuclear magnetic moments
will be examined.


The starting point of the RMF theory is the standard effective
Lagrangian density constructed with the degrees of freedom
associated with nucleon field~($\psi$), two isoscalar meson fields
~($\sigma$ and $\omega_\mu$), isovector meson field~($\vec\rho_\mu$)
and photon field~($A_\mu$). The equation of motion for a
single-nucleon orbit $\psi_i(\bm{r})$ reads,
\begin{equation}
\label{DiracEq}
\{\mathbf{\alpha}\cdot[\bm{p}- \bm{V} (\bm{r})]
  +\beta M^*(\bm{r})+V_0(\bm{r})\}\psi_i(\bm{r})
  =\epsilon_i\psi_i(\bm{r}),
\end{equation}
where $M^*(\bm{r})$ is defined as $M^*(\bm{r})\equiv
M+g_\sigma\sigma(\bm{r})$, with $M$ referring to the mass of bare
nucleon. The repulsive vector potential is $\displaystyle
V_0(\bm{r})=g_\omega\omega_0(\bm{r})+g_\rho\tau_3\rho_0(\bm{r})+e\frac{1-\tau_3}{2}A_0(\bm{r})$,
where $g_i(i=\sigma,\omega,\rho)$ are the coupling strengthes of
nucleon with mesons. The time-odd fields $\bm{V}(\bm{r})$ are
naturally given by the space-like components of vector fields,
$\mbox{\boldmath$V$}(\br) = g_{\omega}\bm{\omega} ({\br})$, where
the space components of $\rho$-meson field $\bm{\rho}(\br)$ and
Coulomb field $\mbox{\boldmath$A$}(\br)$ are neglected since they
turn out to be small compared with $\bm{\omega}(\br)$ field in light
nuclei~\cite{Hofmann1988}.

The non-vanishing time-odd fields in Eq.(\ref{DiracEq}) give rise to
splitting between pairwise time-reversal states $\psi_{i}$ and
$\psi_{\overline i}(\equiv\hat T\psi_{i})$ and also the
non-vanishing current in the core, where $\hat T$ is the
time-reversal operator. Each Dirac spinor $\psi_i(\bm{r})$ and meson
fields are expanded in terms of a set of isotropic harmonic
oscillator basis in cylindrical coordinates with 16 major
shells~\cite{Gambhir1990,Ring1997}. The pairing correlations for
these double-closed shell nuclei plus or minus one nucleon are
neglected due to the quenching effect from unpaired valence nucleon.
More details about solving the Dirac equation with time-odd fields
can be found in Refs.~\cite{Li2009,Li2009a}.

The effective electromagnetic current operator used to describe the
nuclear magnetic moment is given by~\cite{Furnstahl1989,Yao2006}
\begin{equation}
 \hat{J}^\mu(x) =
                   \bar{\psi}(x)\gamma^\mu\frac{1-\tau_3}{2}\psi(x)+\frac{\kappa}{2M}\partial_\nu
                   [\bar{\psi}(x)\sigma^{\mu\nu}\psi(x)],
\end{equation}
where $\sigma^{\mu\nu}=\frac{i} {2} [\gamma^\mu,\gamma^\nu]$, and
$\kappa$ is the free anomalous gyromagnetic ratio of the nucleon,
$\kappa_p=1.793$ and $\kappa_n=-1.913$. The nuclear dipole magnetic
moment is determined by
\begin{eqnarray}\label{magnetic-moment}
  \mbox{\boldmath{$\mu$}}
            &=& \frac{1}{2}\int d^3r
                \br\times\langle g.s. \vert \hat\bj(\br)\vert g.s. \rangle,
\end{eqnarray}
where $\hat\bj(\br)$ is the operator of space-like components of the
effective electromagnetic current.

\begin{figure}[h]
\includegraphics[width=10cm]{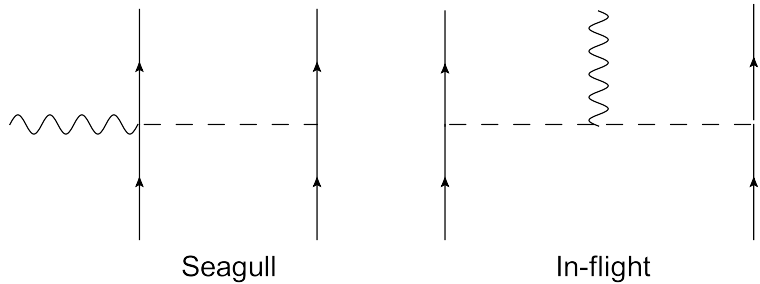}
\vspace{-0.5em} \caption{\label{fig:1} Diagrams of the one-pion
exchange current: seagull (left) and in-flight (right).}
\end{figure}

In addition, for isovector magnetic moment, the one-pion exchange
current correction should be taken into account. Although there is
no explicit pion meson in the RMF theory, it is possible to study
the MEC corrections due to the virtual pion exchange between two
nucleons, which is given by the Feynman diagrams in
Fig.~\ref{fig:1}.

The one-pion exchange current contributions to magnetic moments can
thus be obtained as~\cite{Morse1990},
\begin{eqnarray}\label{magnetic moment-MEC}
    \mbox{\boldmath{$\mu$}}_{\mathrm{MEC}}
    &=& \frac{1}{2}\int d \br\,\br\times
    \langle g.s.|\hat\bj^{\mathrm{seagull}}(\br)
    +\hat\bj^{\mathrm{in\mbox{-}flight}}(\br)|g.s.\rangle,
\end{eqnarray}
where the corresponding one-pion exchange currents
$\hat\bj^{\mathrm{seagull}}(\br)$ and
$\hat\bj^{\mathrm{in\mbox{-}flight}}(\br)$ are respectively,
\begin{eqnarray}
    \hat\bj^{\mathrm{seagull}}(\br)
    &=&-\frac{8ef^2_{\pi}M}{m^2_\pi} \int d \svec x\,
        \bar{\psi}_p(\svec r) {\bm\gamma}\gamma_5\psi_n(\svec r)
         D_\pi(\svec r,\svec x)
    \bar{\psi}_n(\svec y)\frac{M^*}{M}\gamma_5\psi_p(\svec x),\\
    \hat\bj^{\mathrm{in\mbox{-}flight}}(\br)
    &=&-\frac{16ief^2_{\pi}M^2}{m_\pi^2} \int d\svec x d\svec y \bar{\psi}_p(\svec x)\frac{M^*}{M}\gamma_5\psi_n(\svec x)
        D_\pi(\svec x,\svec r){\bm\nabla}_{\svec r}
        D_\pi(\svec r,\svec y)\bar{\psi}_n(\svec y)\frac{M^*}{M}\gamma_5\psi_p(\svec
        y),\nonumber\\
\end{eqnarray}
with the $\pi$-nucleon coupling constant $f_\pi=1$ and the pion mass
$m_\pi=138$ MeV.  The pion propagator in r-space is given by
$D_\pi(\svec x,\svec r)=\dfrac{1}{4\pi}\dfrac{e^{-m_\pi|\svec
x-\svec r|}} {|\svec x-\svec r|}$.

The magnetic moments of double-closed shell nuclei plus or minus one
nucleon with $A=15, 17, 39$ and 41 are studied in the RMF theory
using PK1 effective interaction~\cite{Long2004}, which includes the
self-couplings of $\sigma$ and $\omega$ mesons.

The magnetic moments in Eq.(\ref{magnetic-moment}) will be
calculated using Dirac spinors $\psi_i$ from the axially deformed
RMF calculations with space-like components of vector meson field.
As small deformation in these nuclei, the one-pion exchange current
contributions to magnetic moments in Eq.(\ref{magnetic moment-MEC})
are calculated using the spherical Dirac spinors of corresponding
double-closed shell nucleus as done in Ref.~\cite{Morse1990}.


\begin{table}[h!]\tabcolsep=5pt
\caption{\label{tab:mec}The one-pion exchange current corrections to
the isovector magnetic moments obtained from RMF calculations using
PK1 effective interaction, in comparison with the Linear
RMF~\cite{Morse1990} and non-relativistic
results~\cite{Chemtob1969,Hyuga1980,Towner1983,Ito1987} (see text
for details).}
\begin{tabular}{cccccccc}
  \hline\hline
\multirow{2}{*}{~~A~~} &\multicolumn{4}{c}{Non-relativistic}&
&\multicolumn{2}{c}{Relativistic}
\\ \cline{2-5}\cline{7-8}
  &\cite{Chemtob1969} & \cite{Hyuga1980}& \cite{Towner1983}&     \cite{Ito1987} && \cite{Morse1990} & This
work
\\ \hline
15 & 0.127 & 0.116& 0.092&  0.111 && 0.102 & 0.091 \\
17 & 0.084 & 0.093& 0.065&  0.092 && 0.151 & 0.092  \\
39 & 0.204 & 0.199& 0.149&  0.184 && 0.174 & 0.190  \\
41 & 0.195 & 0.201& 0.115&  0.180 && 0.270 & 0.184  \\
  \hline
\end{tabular}
\end{table}

The one-pion exchange current corrections to the isovector magnetic
moments obtained from RMF calculations using PK1 are compared in
Table~\ref{tab:mec} with linear RMF calculations~\cite{Morse1990}
using L3~\cite{Lee1986} and non-relativistic
calculations~\cite{Chemtob1969,Hyuga1980,Towner1983,Ito1987}. It is
shown that the obtained corrections to the isovector magnetic
moments in this work are in reasonable agreement with other
calculations. As noted in Ref.~\cite{Morse1990}, the differences
between the various calculations presented in Table~\ref{tab:mec}
are most likely due to relatively small changes in the balance of
contributions from seagull and in-flight diagrams rather than any
fundamental differences in the models used. Other nonlinear
effective interactions are also used to calculate the one-pion
exchange current corrections, and similar results are obtained as
those given by PK1.

\begin{table}[h!]\tabcolsep=5pt
\caption{\label{tab:isoscalar}Isoscalar magnetic moments obtained
from RMF calculations using PK1 effective interaction, in comparison
with the corresponding data, Schmidt value, previous relativistic
result~\cite{Morse1990} and non-relativistic
results~\cite{Towner1987,Arima1987}(see text for details).}
\begin{tabular}{cccccccc}
  \hline\hline
\multirow{2}{*}{~~A~~} & \multirow{2}{*}{Exp.}
&\multicolumn{3}{c}{Non-relativistic}&
&\multicolumn{2}{c}{Relativistic}
\\ \cline{3-5}\cline{7-8}
 & &Schmidt &~\cite{Towner1987}~&~\cite{Arima1987} && QHD+MEC~\cite{Morse1990} & RMF+MEC
\\ \hline
15  &0.218 &0.187  &0.228 &0.233 & & 0.200(0.199$+$0.001)& $0.216(0.216+0.000)$\\
17  &1.414 &1.440  &1.410 &1.435 & & 1.42~(1.43~$-$0.011)& $1.467(1.469-0.002)$ \\
39  &0.706 &0.636  &0.706 &0.735 & & 0.659(0.660$-$0.001)& $0.707(0.707+0.000)$ \\
41  &1.918 &1.940  &1.893 &1.944 & & 1.93~(1.94~$-$0.007)& $1.988(1.991-0.003)$ \\
  \hline
\end{tabular}
\end{table}
In Table~\ref{tab:isoscalar}, the isoscalar magnetic moments and
corresponding pion exchange current corrections obtained from RMF
calculations using PK1 are presented in comparison with the
corresponding data, Schmidt value, previous relativistic
result~\cite{Morse1990} and non-relativistic
results~\cite{Towner1987,Arima1987}. The isoscalar magnetic moments
obtained from deformed RMF theory with space-like components of
vector meson are labeled as RMF and corresponding one-pion exchange
current corrections calculated similarly as in Ref.~\cite{Morse1990}
are labeled as MEC.

The isoscalar magnetic moment in Ref.~\cite{Morse1990} consists of
two parts, i.e., the QHD calculations taken from
Ref.~\cite{Furnstahl1989} and the additional one-pion exchange
current corrections calculated with L3 effective interaction.

For the non-relativistic calculations in
Refs.~\cite{Towner1987,Arima1987}, the harmonic oscillator wave
functions are used for single-particle states and one-boson-exchange
potential~\cite{Towner1987} and Hamada-Johnston
potential~\cite{Arima1987} were respectively employed for the
residual interaction. For the corrections to magnetic moments, the
second-order core polarization, MEC, and the crossing term between
MEC and core polarization have been included. For the MEC
corrections, the $\Delta$ isobar current as well as the exchange
current of the mesons $\pi$, $\sigma$, $\omega$, and $\rho$ have
been taken into account.

It is shown that all calculated results in Table~\ref{tab:isoscalar}
are in good agreement with data, and same as the previous
relativistic~\cite{Morse1990} and non-relativistic
calculations~\cite{Towner1987,Arima1987}, the MEC corrections to
isoscalar moments in present calculations are negligible. For the
mirror nuclei with double-closed shell plus or minus one nucleon,
the MEC corrections to isoscalar moments reflect the violation of
isospin symmetry in wave functions. With the small MEC corrections
to isoscalar moments here, it is easy to understand the excellent
description of the isoscalar magnetic moments in deformed RMF theory
with space-like components of vector meson in
Refs.~\cite{Hofmann1988,Yao2006}.

\begin{table}[t!h]\tabcolsep=5pt
\caption{\label{tab:isovector} Same as Table~\ref{tab:isoscalar},
but for the isovector magnetic moments.}
\begin{tabular}{cccccccc}
  \hline\hline
\multirow{2}{*}{~~A~~} & \multirow{2}{*}{Exp.}
&\multicolumn{3}{c}{Non-relativistic}&
&\multicolumn{2}{c}{Relativistic}
\\ \cline{3-5}\cline{7-8}
 & &Schmidt &~\cite{Towner1987}~&~\cite{Arima1987} && QHD+MEC~\cite{Morse1990} & RMF+MEC
\\ \hline
15  &$-0.501$      &$-0.451$      &$-0.456$      &$-0.508$       & &$-0.347(-0.449+0.102)$           & $-0.339(-0.430+0.091)$          \\
17  &$\;\;\,3.308$ &$\;\;\,3.353$ &$\;\;\,3.281$ &$\;\;\,3.306$  & &$\;\;\,3.61\;\;(\;\;\,3.46\;\;+0.151)$ & $\;\;\,3.576(\;\;\,3.483+0.092)$ \\
39  &$-0.316$      &$-0.512$      &$-0.286$      &$-0.481$       & &$-0.106(-0.280+0.174)$           & $-0.115(-0.305+0.190)$           \\
41  &$\;\;\,3.512$ &$\;\;\,3.853$ &$\;\;\,3.803$ &$\;\;\,3.729$  & &$\;\;\,4.41\;\;(\;\;\,4.14\;\;+0.270)$ & $\;\;\,4.322(\;\;\,4.138+0.184)$ \\
  \hline
\end{tabular}
\end{table}

In Table~\ref{tab:isovector}, the isovector magnetic moments and
corresponding pion exchange current corrections in RMF calculations
using PK1 are compared with the data, Schmidt value, previous
relativistic~\cite{Morse1990} and non-relativistic
results~\cite{Towner1987,Arima1987}.

It is shown that the pion exchange current gives a significant
positive correction to isovector magnetic moments, which is
consistent with the calculations in Ref.~\cite{Morse1990} as well as
most non-relativistic calculations~\cite{Towner1987,Arima1987}.
However, compared with the case for the isoscalar magnetic moments,
the results of relativistic calculations deviate much more from data
explicitly, namely, this positive contribution is not welcome to
improve the agreement with data. Such a phenomenon is also found
from RMF calculations with other effective interactions. Therefore,
the RMF theory with one-pion exchange current corrections could not
improve the description of isovector magnetic moment for the
concerned nuclei.

In the future relativistic investigation, the other effects due to
the second-order core polarization, the $\Delta$ isobar current,
exchange current corrections due to other mesons, and the crossing
term between MEC and core polarization should be taken into account,
as noted already in the non-relativistic
calculations~\cite{Towner1987,Arima1987}.

In summary, the one-pion exchange current corrections to the
isoscalar and isovector magnetic moments have been studied in the
RMF theory with PK1 effective interaction and compared with previous
relativistic and non-relativistic results. It has been found that
the one-pion exchange current gives a negligible contribution to the
isoscalar magnetic moments but a significant correction to the
isovector ones. However, the one-pion exchange current doesn't
improve the description of nuclear isovector magnetic moments for
the concerned nuclei. In the future investigation, similar as the
non-relativistic cases~\cite{Towner1987,Arima1987}, the second-order
core polarization effects, the $\Delta$ isobar current, crossing
term between MEC and core polarization, and exchange current
corrections due to other mesons should be taken into account. In
addition, the correction due to the restoration of the rotational
symmetry~\cite{Yao2009} may play a role as well. The investigation
towards these directions is in progress.


\begin{acknowledgments}
We would like to thank W. Bentz for his careful reading of the
manuscript and comments. This work is partly supported by Major
State Basic Research Developing Program 2007CB815000, the National
Natural Science Foundation of China under Grant Nos. 10775004,
10720003, 10947013, 10975008, and 10975007, as well as the Southwest
University Initial Research Foundation Grant to Doctor No.
SWU109011.
\end{acknowledgments}

\end{document}